\documentclass[11pt]{article}
\usepackage{moriond,epsfig}

\def\Journal#1#2#3#4{{#1} {\bf #2}, #3 (#4)}

\def\NPB{{\em Nucl. Phys.} B}
\def\PLB{{\em Phys. Lett.}  B}

\def\PRD{{\em Phys. Rev.} D}

\def\be{\begin{equation}}
\def\ee{\end{equation}}
\def\bea{\begin{eqnarray}}
\def\eea{\end{eqnarray}}

\begin{document}
\vspace*{4cm}
\title{SEARCHES FOR GAUGE MEDIATED SUSY BREAKING AT LEP}

\author{ K. KLEIN }

\address{University of Heidelberg, Physikalisches Institut, \\ 
Philosophenweg 12, 69120 Heidelberg, Germany}

\maketitle\abstracts{Searches for neutralinos and sleptons with arbitrary 
lifetimes, as predicted in the framework of the Gauge Mediated Supersymmetry 
Breaking (GMSB) model, have been performed by the four LEP collaborations. No 
evidence for these particles has been found in the data recorded at 
center--of--mass energies up to $\sqrt{s}=209\,$GeV. Therefore constraints on 
the production 
cross--sections and particle masses as well as interpretations in the 
framework of the GMSB model are presented.}

\section{Introduction}\label{sec:intro}
In supersymmetry (SUSY)~\cite{susy} for every Standard Model (SM) particle 
a supersymmetric ``partner'', differing in spin by half a unit, is predicted. 
If SUSY were an exact symmetry, the supersymmetric particles would be 
degenerate in mass with their Standard Model partners. However, since no SUSY 
particles have been observed so far, it can be 
deduced that the SUSY partners must be much heavier than (most of) the 
Standard Model particles, and therefore SUSY must be a broken symmetry. 
Several mechanisms for SUSY breaking have been considered. One approach is 
that SUSY is broken in a ``hidden'' sector, which couples to 
the visible sector containing the SM and SUSY particles via gravitational 
interactions (gravity mediation). Another possibility is that the hidden 
sector couples only to a ``messenger sector'', which in turn couples via the 
gauge interactions to the visible sector of the SM and SUSY 
particles~\cite{theo1,theo2,theo3}. This 
model is called the Gauge Mediated Supersymmetry Breaking (GMSB) model. In 
its minimal version six new parameters are introduced in addition to the SM 
parameters, usually chosen to be the SUSY breaking scale, $\sqrt{F}$, the 
messenger scale, $M$, the messenger index, $N$, the 
ratio of the vacuum expectation values of the two 
Higgs doublets, $\tan{\beta}$, the sign of the Higgs sector mixing 
parameter, sign($\mu$), and the mass scale $\Lambda$, which determines the 
SUSY particle masses at the messenger scale.

In supersymmetric theories the phenomenology depends crucially on the 
nature of the lightest and next--to--lightest supersymmetric particles, the 
LSP and the NLSP. The mass of the gravitino $\tilde{\rm G}$, the SUSY partner 
of the graviton, is determined by the SUSY breaking scale $\sqrt{F}$. In GMSB 
models the SUSY breaking scale is typically low 
(${\mathcal O}(100\,\rm TeV)$), which leads to a light gravitino and 
makes the gravitino the LSP. Depending on the choice of 
the other model parameters, the NLSP can either be the lightest neutralino, 
$\tilde{\chi}_1^0$, which is a mixture of the SUSY partners of the $\gamma$, 
Z$^0$ and the neutral Higgs bosons, or the lightest SUSY partner of the 
leptons, a slepton $\tilde{\ell}_R$. In the latter case two scenarios are 
distinguished: the stau NLSP scenario, in which the lighter stau, 
$\tilde{\tau}_1$, is lighter than the other sleptons and is the sole NLSP, 
and the slepton co--NLSP scenario, in which all sleptons are degenerate in 
mass and act as ``co--NLSPs''. 

In GMSB models the proper decay length $L$ of the NLSP depends on $\sqrt{F}$: 
\begin{equation}
L =\frac{0.01}{\kappa_{\gamma}}\left(\frac{100\,\mathrm{GeV}}
m\right)^5\left(\frac{\sqrt{F}}{100\,\mathrm{TeV}}\right)^4\mathrm{cm}\,,
\label{eq:1}
\end{equation}
where $m$ is the mass of the NLSP and $\kappa_{\gamma}$ is the photino 
component of the neutralino and is equal to one 
for sleptons. Taking into account the range allowed for $\sqrt{F}$, the NLSP 
decay length is basically arbitrary and all possible decay lengths between 
zero and infinity have to be considered. 

In the following, R--parity conservation is assumed, implying that 
SUSY particles are produced only in pairs and that all decay chains 
terminate with SM particles plus the LSP, which is stable. 

The four LEP collaborations, ALEPH, DELPHI, L3 and OPAL, have searched for a 
great variety of topologies expected from GMSB models. Here a selection of 
these results, based mainly on the data 
recorded from 1998--2000 at center--of--mass energies between 189\,GeV and 
209\,GeV, is presented, including the results of a preliminary LEP 
combination. All presented limits are at 95\,\% confidence level (C.L.).

\section{The Neutralino as the NLSP}
The neutralino can be produced directly in pairs, 
$\rm e^+e^-\rightarrow\tilde{\chi}_1^0\tilde{\chi}_1^0$, via the $s$--channel 
with $\gamma/{\rm Z^0}$ exchange, or the $t$--channel with an exchange of a 
selectron. Indirect production is also possible, e.g.~via pair--production of 
sleptons, 
$\rm e^+e^-\rightarrow\tilde{\ell}_R^+\tilde{\ell}_R^-\rightarrow\ell^+\tilde{\chi}_1^0\,\ell^-\tilde{\chi}_1^0$, or pair--production of charginos, 
$\rm e^+e^-\rightarrow\tilde{\chi}_1^+\tilde{\chi}_1^-\rightarrow$$W^{+*}\tilde{\chi}_1^0\,$$W^{-*}\tilde{\chi}_1^0$. If the neutralino is the NLSP, it 
decays to a photon and a gravitino, which escapes detection. Therefore the 
event topology implies photons plus missing energy, possibly with 
additional detector activity from leptons or jets, depending on the 
production channel.

\subsection{Searches for Neutralino Pair--Production}
\begin{figure}
\psfig{figure=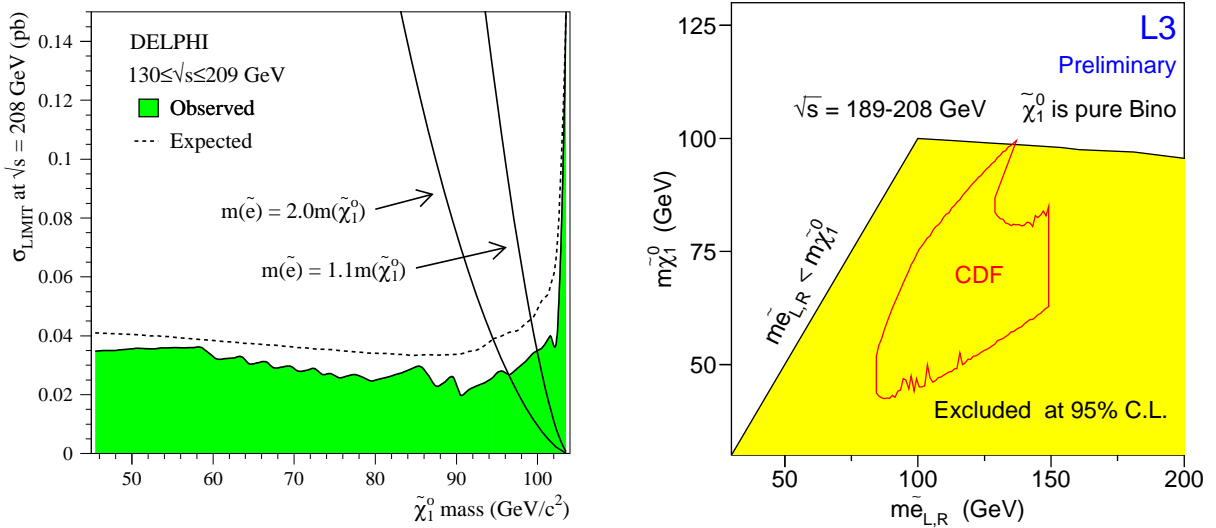,height=7.cm}
\hfill
\psfig{figure=l3_cdflim.epsi,height=7.cm}
\caption{\label{f:neutralino}
Left: Upper limit at 95\,\% C.L. on the cross--section for neutralino 
pair--production for $\sqrt{s}=208$\,GeV from DELPHI (shaded histogram), as a 
function of the neutralino mass. The 
dashed line gives the expected limit, while the theoretical expectations for 
two ratios of the selectron to the neutralino mass are shown as solid lines. 
Right: 95\,\% C.L. excluded region in the selectron -- neutralino mass plane 
from L3, assuming the neutralino to be pure bino. Overlayed is the region 
consistent with the GMSB interpretation of the 
$\rm ee\gamma\gamma\!\not\!{E}_T$ event from CDF.}
\end{figure}

If neutralinos are produced in pairs and decay promptly, two 
high--momentum acoplanar photons are expected in the detector. The dominating 
Standard Model background for this topology is neutrino pair--production with 
two photons from initial state radiation. All four LEP collaborations have 
searched for this topology, but no excesses were 
found~\cite{aleph,delphi2,l3,opal2}. The 
upper limit on the production cross--section from DELPHI~\cite{delphi2}, at 
95\,\% C.L. and for $\sqrt{s}=208\,$GeV, is shown on the left side of 
Figure~\ref{f:neutralino}. Typically, 
cross--sections above 0.04\,pb are excluded. The theoretical cross--section 
depends on the mass of the exchanged selectron. For 
$M(\tilde{\rm e})=2\cdot M(\tilde{\chi}_1^0)$ a lower neutralino mass limit 
of 96\,GeV is found at 95\,\% C.L., while for 
$M(\tilde{\rm e})=1.1\cdot M(\tilde{\chi}_1^0)$ the limit increases to 
100\,GeV. For the determination of these limits, the right-- and left--handed 
selectrons are assumed to be mass--degenerate and the neutralino is assumed 
to be pure bino.

On the right side of Figure~\ref{f:neutralino} the excluded region in the 
selectron -- neutralino mass plane from the L3 collaboration is 
shown~\cite{l3}, for a pure bino 
neutralino. The region consistent with the GMSB interpretation of the 
${\rm ee}\gamma\gamma\!\not{\!\!E_T}$ event observed by CDF~\cite{cdf} 
($\mathrm{\tilde{e}^+\tilde{e}^-\rightarrow \tilde{\chi}_1^0\tilde{\chi}_1^0 
e^+e^- \rightarrow \gamma\gamma\tilde{G}\tilde{G}e^+e^-}$) is completely 
excluded at 95\,\% C.L.\,. 

\setlength{\unitlength}{1.cm}

If the lightest neutralino has a lifetime such that it does not decay 
promptly but before reaching the electromagnetic calorimeter, the 
experimental topology is photons 
that do not point to the primary event vertex. Both ALEPH and DELPHI have 
searched for such non--pointing photons, requiring at least one photon with a 
reconstructed impact parameter larger than 40\,cm. ALEPH selects two 
candidates in the data set recorded at $\sqrt{s}=189-209$\,GeV, with 1.0 
events expected from Standard Model sources~\cite{aleph}, while DELPHI, using 
the data collected at $\sqrt{s}=130-209$\,GeV, selects 16 candidates with 
14.6 events expected~\cite{delphi2}. 
DELPHI reports 95\,\% C.L. cross--section limits of the order of 0.4\,pb for 
mean neutralino decay lengths of approximately 2--20\,m, taking into account 
the data collected at $\sqrt{s}=192-209$\,GeV.

\subsection{Searches for Indirect Neutralino Production}
If the neutralino is stable or both neutralinos decay outside the detector, 
direct neutralino pair--production is invisible. The indirect 
production channels, however, can be used to search for very long--lived 
neutralinos. Utilising the searches for slepton and chargino pair--production 
in gravity mediated models, where the neutralino is the NLSP, the ALEPH 
collaboration reports a lower neutralino mass limit of 54\,GeV at 
95\,\% C.L., independent of the neutralino lifetime~\cite{aleph}. 

\section{The Slepton as the NLSP}
If the slepton is the NLSP, it decays to a lepton of the same flavour and 
a gravitino, leading to final states with leptons and missing energy. Again, 
different production channels are possible. The most important one is direct 
slepton pair--production, 
${\rm e^+e^-}\rightarrow\tilde{\ell}_R^+\tilde{\ell}_R^-$, which proceeds via 
the $s-$channel and, for selectrons only, also the $t-$channel with neutralino 
exchange. Indirect production is possible at LEP2 via pair--production of 
neutralinos,
 ${\rm e^+e^-}\rightarrow\tilde{\chi}_1^0\tilde{\chi}_1^0\rightarrow\tilde{\ell}_R^{\pm}\ell^{\mp}\,\tilde{\ell}_R^{'\pm}\ell'^{\mp}$, and charginos, 
${\rm e^+e^-}\rightarrow\tilde{\chi}_1^+\tilde{\chi}_1^-\rightarrow\tilde{\ell}^+_R\,\nu\,\tilde{\ell}^-_R\,\bar\nu$. Finally, 
if only the stau is the NLSP, it can be produced indirectly via 
pair--production of selectrons or smuons: 
${\rm e^+e^-}\rightarrow\tilde{\ell}_R^+\tilde{\ell}_R^-\rightarrow
\ell^+\tilde{\chi}_1^0\,\ell^-\,\tilde{\chi}_1^0\rightarrow
\ell^+\tilde{\tau}^{\pm}_1\tau^{\mp}\,\ell^-\tilde{\tau}^{\pm}_1\tau^{\mp}$, 
with $\tilde{\ell}_R=\tilde{\rm e}_R,\,\tilde{\mu}_R$. 
Results will be reported only on the first three channels.

\subsection{Searches for Slepton Pair--Production}
If sleptons are pair--produced and decay promptly, the experimental 
topology is two high--energetic acoplanar lepton tracks. The same topology is 
expected in gravity mediated models, where the slepton decays to a lepton and 
a stable neutralino, if the neutralino is assumed to be very light. Thus the 
searches for slepton pair--production in the context of the minimal 
Supergravity model are used. All four LEP collaborations have searched for 
acoplanar lepton events, but no significant excesses were 
observed~\cite{aleph,l3,opal2,delphi1}. The 
LEP--combined upper cross--section limits at 95\,\% C.L., including the data 
with $\sqrt{s}=183-209\,$GeV, are below 0.05\,pb for smuons and below 
0.18\,pb for staus~\cite{lepsusyacop}.


If the sleptons have intermediate lifetimes and decay either before or inside 
the tracking devices, the topologies of tracks with large impact parameters 
and tracks with kinks are expected. All four LEP collaborations have searched 
for these interesting topologies, but observe no excess over the background 
expectation~\cite{aleph,l3,delphi1,opal}. 

Finally, if the sleptons are stable or both sleptons decay outside the 
tracking devices, they will distinguish themselves from Standard Model 
particles by their anomalously high or low specific energy loss, d$E$/d$x$, 
due to their large mass. 
All LEP collaborations have performed searches for this almost 
background--free topology~\cite{aleph,l3,delphi1,opal}. No hint for the 
production of such long--lived sleptons was found.

\begin{figure}
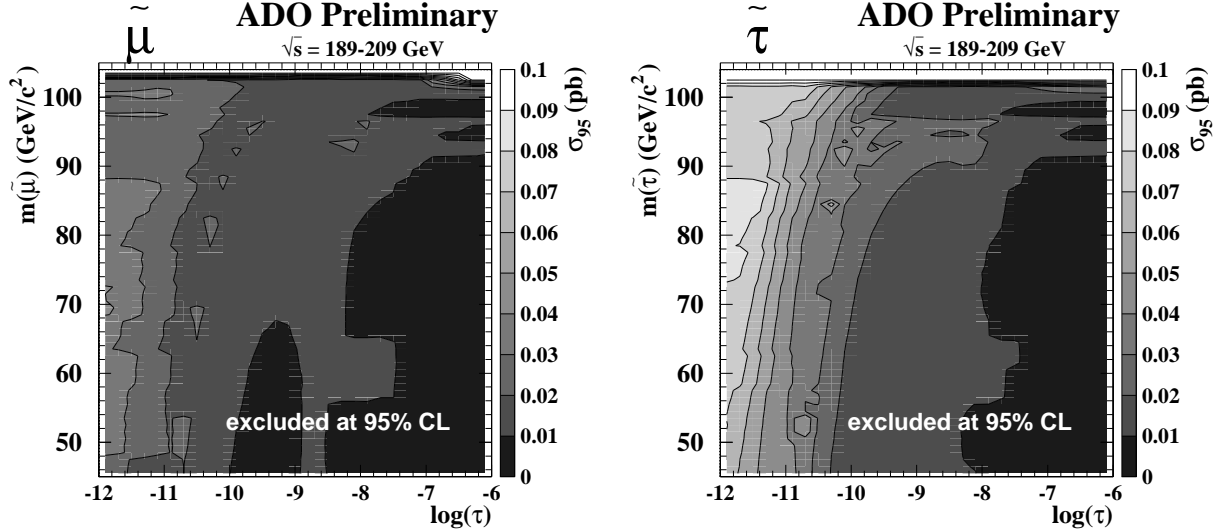

\psfig{figure=lep_smuon_xsec_bw.epsi,height=7.cm}
\hfill
\psfig{figure=lep_stau_xsec_bw.epsi,height=7.cm}
\caption{
\label{f:slepton}
Contours of the upper limits on the production cross--sections at 
95\,\% C.L. for smuons (left) and staus (right) in the slepton mass -- 
lifetime plane, combining the results from ALEPH, DELPHI and OPAL.}
\end{figure}

\begin{figure}
\hfill
\psfig{figure=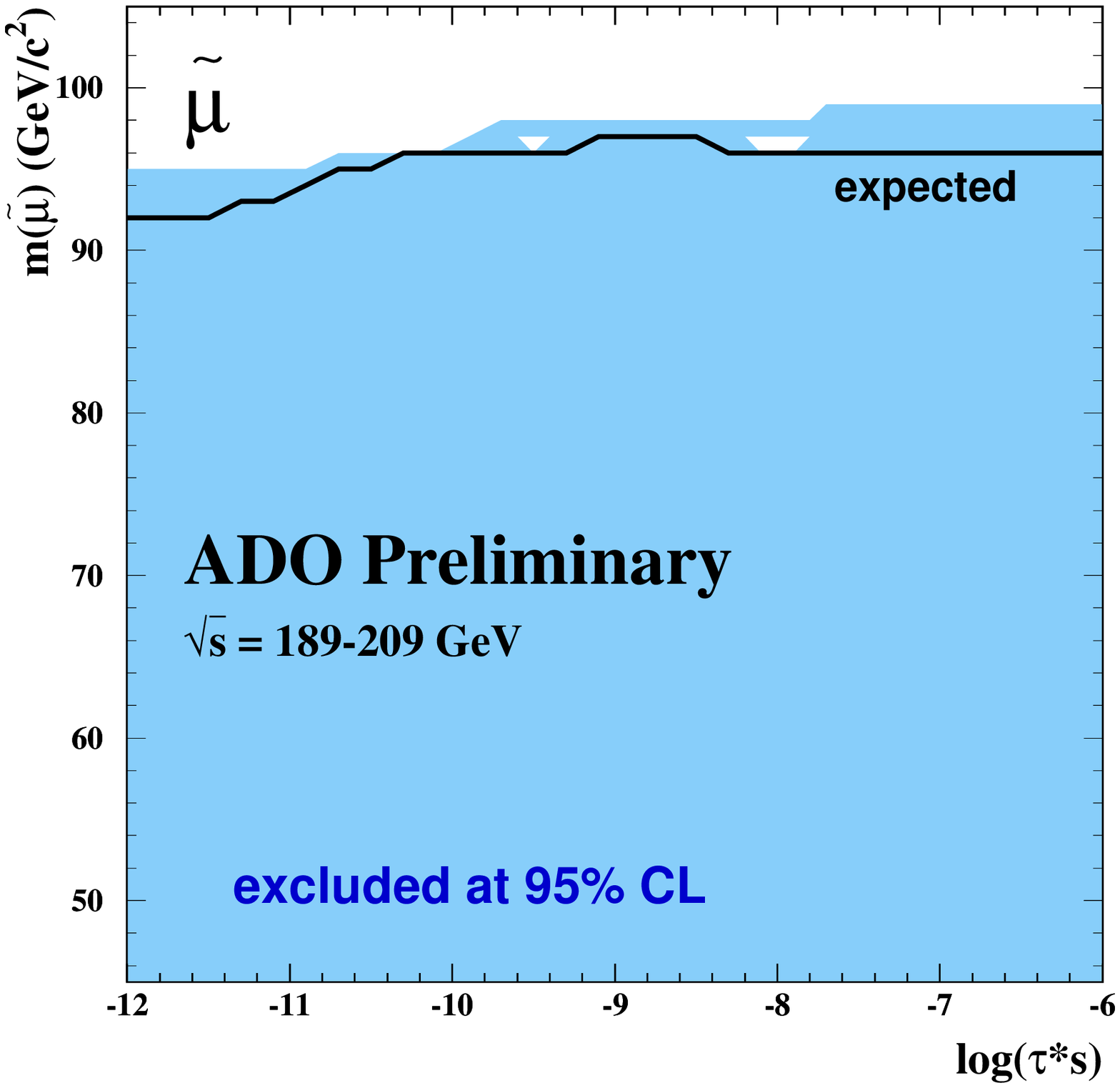,height=6.5cm}
\hfill
\psfig{figure=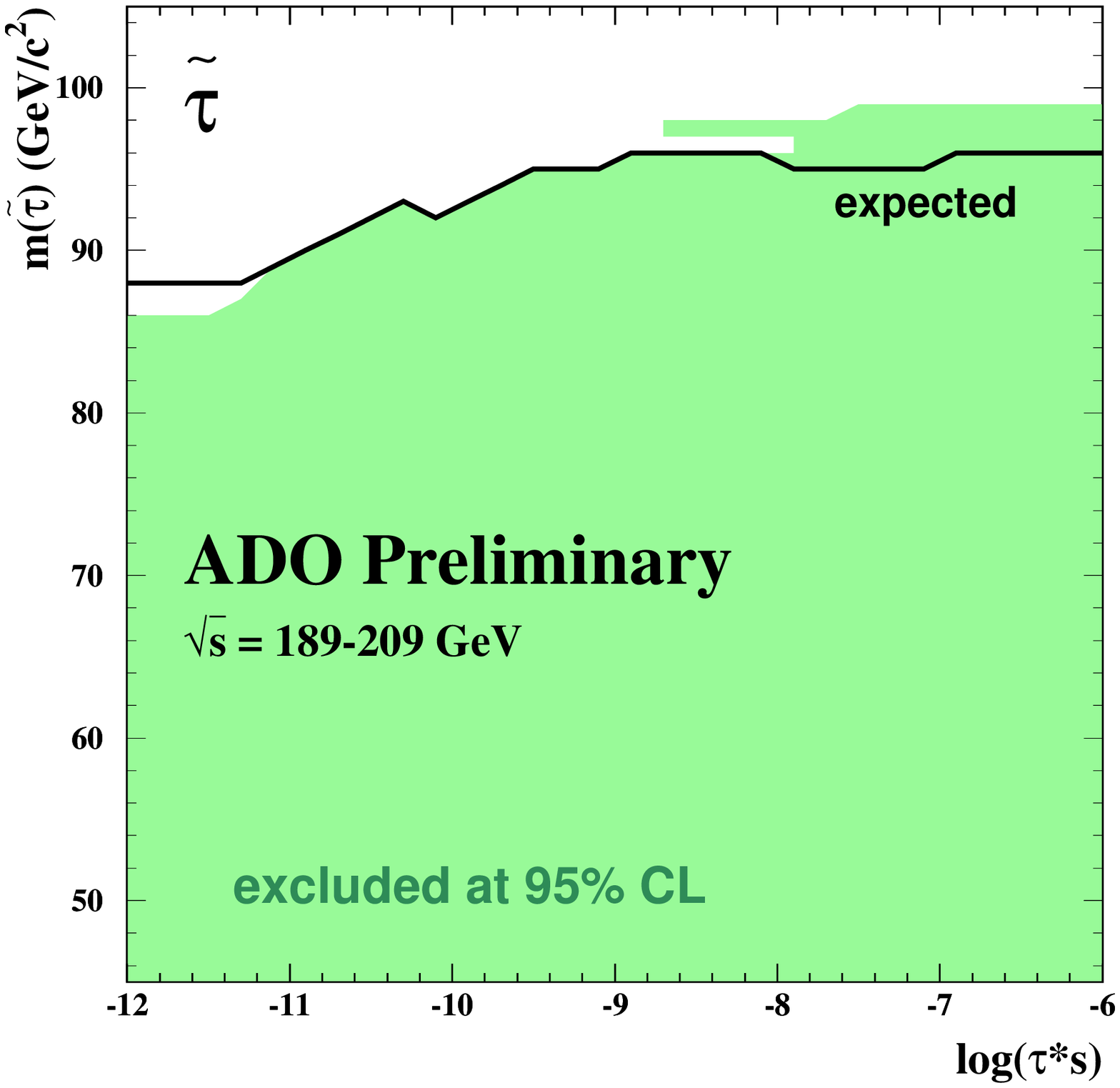,height=6.5cm}
\hfill
\caption{
\label{f:sleptonmass}
Excluded regions at 95\,\% C.L. in the slepton mass -- lifetime plane for 
smuons (left) and staus (right), combining the results from ALEPH, DELPHI and 
OPAL. The expected mass limits are shown as solid lines.}
\end{figure}

Combining the searches for sleptons of all lifetimes, cross--section and mass 
limits in the slepton mass -- lifetime plane are obtained. A preliminary LEP 
combination, including the results of ALEPH, DELPHI and OPAL at 
center--of--mass energies of $\sqrt{s}=189-209\,$GeV~\footnote{The L3 
collaboration presents results on slepton pair--production only for a 
subset of the available data ($\sqrt{s}=200\,$GeV).}, was 
performed~\cite{lepsusygmsb}. 
The resulting cross--section limits at $\sqrt{s}=206\,$GeV, as a function of 
the lifetime, are 
shown in Figure~\ref{f:slepton} for smuons (left) and staus (right). 
Cross--sections above 0.07\,pb, 0.04\,pb and 0.09\,pb are excluded for 
selectrons, smuons and staus, respectively. Figure~\ref{f:sleptonmass} 
shows the lower limits on the slepton masses, as a function of the slepton 
lifetime, for smuons (left) and staus (right). The lowest limit is found for 
small lifetimes, due to irreducible background from leptonic W decays to 
the acoplanar lepton topology. For $N\le 5$, slepton mass 
limits of $M(\tilde{\rm e}_R)>66.0\,$GeV, $M(\tilde{\mu}_R)>95.2\,$GeV and 
$M(\tilde{\tau}_1)>86.1\,$GeV were obtained, independent on the slepton 
lifetime. For these limits, the theoretical cross--sections were 
taken from a scan based on the framework of~\cite{theo3}.

\setlength{\unitlength}{1.cm}

\setlength{\unitlength}{1.cm}

\subsection{Searches for Neutralino Pair--Production}
If the slepton is the NLSP and is produced via pair--production of 
neutralinos, the topology is four leptons, of which two might originate from 
the decays of long--lived sleptons, plus missing energy. ALEPH~\cite{aleph}, 
DELPHI~\cite{delphi1} and OPAL~\cite{opal} have searched for neutralino 
pair--production with promptly decaying sleptons, and ALEPH and OPAL consider 
in addition the case that the tracks of two of 
the decay leptons might have large impact parameters or kinks. All three 
collaborations report good agreement between the number of candidates and 
the number of events expected from SM sources. By combining this channel with 
the acoplanar lepton and acoplanar photon searches, ALEPH obtains a lower 
mass limit for the neutralino of 92\,GeV for the case of zero NLSP lifetime.

\subsection{Searches for Chargino Pair--Production}
DELPHI has searched for chargino pair--production with a slepton 
NLSP~\cite{delphi1}, taking 
into account all slepton lifetimes. Since no evidence for the production of 
charginos has been observed, DELPHI sets chargino mass limits of 100\,GeV and 
96\,GeV at 95\,\% C.L. in the stau NLSP scenario and the slepton--co NLSP 
scenario, respectively, independent of the slepton lifetime. 

\section{Model--Dependent Interpretations}
\begin{figure}
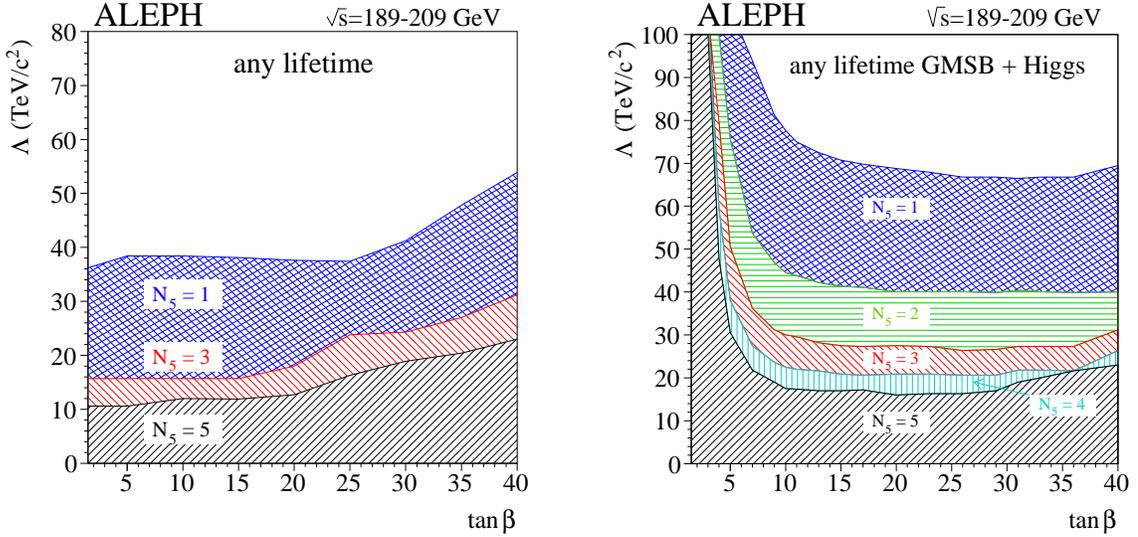

\hfill
\psfig{figure=aleph_lambtanb_all_mod.epsi,height=7.cm}
\hfill
\psfig{figure=aleph_lambtanb_higgs.epsi,height=7.cm}
\hfill
\caption{
\label{f:interpretation}
Left: Excluded regions in the $\Lambda-\tan{\beta}$ plane for different 
values of $N$, as derived by ALEPH for all NLSP lifetimes. Right: 
larger regions can be excluded if the ALEPH results of searches for neutral 
Higgs bosons are taken into account.}
\end{figure}

To interpret the experimental results in the framework of the GMSB model, the 
ALEPH, DELPHI and OPAL collaborations each perform a scan over the GMSB 
parameters, and exclude points in this GMSB parameter space. On the left side 
of Figure~\ref{f:interpretation} the regions in 
the $\Lambda-\tan{\beta}$ plane excluded by ALEPH~\cite{aleph} are shown for 
different 
values of $N$ and for all NLSP lifetimes. A limit on the parameter $\Lambda$, 
which determines the SUSY particles masses at the messenger scale, of 
$\Lambda>10\,$TeV 
has been found for $N\le 5$. When the ALEPH results on neutral Higgs boson 
searches~\cite{alephhiggs} are taken into account, larger regions in the 
parameter space can be excluded, as shown in the plot on the right side of 
Figure~\ref{f:interpretation}, and the lower limit on $\Lambda$ increases to 
16\,TeV for $N\le 5$. Using the condition 
$\Lambda\le\sqrt{F}$ and the relation between the gravitino mass and 
$\sqrt{F}$, these limits can be translated into lower limits on the gravitino 
mass of $M(\tilde{\rm G})>0.024$\,eV and $M(\tilde{\rm G})>0.061$\,eV, 
respectively. The DELPHI collaboration reports a limit of $\Lambda>17.5$\,TeV 
for $N\le 4$ and the case of a negligible NLSP lifetime~\cite{delphi1}.

\setlength{\unitlength}{1.cm}

\section{Conclusions}
The four LEP collaborations have performed searches for neutralinos and 
sleptons with arbitrary lifetimes, as predicted in the framework of the GMSB 
model. Since no 
evidence for the production of such particles has been observed, 
cross--section and mass limits have been obtained using the data recorded at 
center--of--mass energies up to 209\,GeV, and the experimental results were 
used to constrain the GMSB parameter space.

\end{document}